\documentclass[pra,aps,twocolumn,superscriptaddress]{revtex4-1}
\usepackage{graphicx}
\usepackage{multirow}
\usepackage{color}
\usepackage{ulem}

\newcommand{\bra}[1] {\left\langle #1 \right|}
\newcommand{\ket}[1] {\left| #1 \right\rangle}

\begin{document}

\title{Characteristic spectra of circuit quantum electrodynamics systems from the ultrastrong- to the deep- strong-coupling regime}

\author{F. Yoshihara}
\email{fumiki@nict.go.jp}
\affiliation{National Institute of Information and Communications Technology, 4-2-1, Nukuikitamachi, Koganei, Tokyo 184-8795, Japan}
\author{T. Fuse}
\affiliation{National Institute of Information and Communications Technology, 4-2-1, Nukuikitamachi, Koganei, Tokyo 184-8795, Japan}
\author{S. Ashhab}
\affiliation{Qatar Environment and Energy Research Institute, Hamad Bin Khalifa University, Qatar Foundation, Doha, Qatar}
\author{K. Kakuyanagi}
\affiliation{NTT Basic Research Laboratories, NTT Corporation, 3-1 Morinosato-Wakamiya, Atsugi, Kanagawa 243-0198, Japan}
\author{S. Saito}
\affiliation{NTT Basic Research Laboratories, NTT Corporation, 3-1 Morinosato-Wakamiya, Atsugi, Kanagawa 243-0198, Japan}
\author{K. Semba}
\affiliation{National Institute of Information and Communications Technology, 4-2-1, Nukuikitamachi, Koganei, Tokyo 184-8795, Japan}

\date{\today}

\begin{abstract}
We report on spectra of circuit-quantum-electrodynamics (QED) systems in an intermediate regime that lies between the ultrastrong and deep-strong-coupling regimes, which have been reported previously in the literature.
Our experimental results, along with numerical simulations, demonstrate that as the coupling strength increases, the spectrum of a circuit-QED system undergoes multiple qualitative transformations, such that  several coupling regimes
are identified, each with its own unique spectral features.
The different spectral transformations can be related to crossings
between energy level differences and to changes in the symmetries of the energy eigenstates.
These results allow us to use qualitative spectral features to infer certain properties and parameters of the system.
\end{abstract}


\maketitle

\section{Introduction}
Cavity-QED, which describes the interaction between atoms or atom-like emitters and an electromagnetic cavity or more generally any harmonic oscillator, has been an active research area for several decades \cite{Jaynes}. The realization of superconducting circuits where a single qubit is coupled to a superconducting resonator in the field now known as circuit-QED has resulted in remarkable advances in this field.
The strong-coupling regime, where the qubit-oscillator coupling strength $g$ exceeds the relevant decay rates, was realized in 2004~\cite{Chiorescu,Wallraff}.
Several years later,
the ultrastrong-coupling regime was realized~\cite{Niemczyk,FornDiaz2010},
where $g$ was around 10\% of the oscillator's frequency $\omega$
and deviations from the predictions based on the rotating-wave approximation were observed.
Very recently, the deep-strong-coupling regime has also been realized \cite{Yoshihara}.
In this new regime,
$g$, $\omega$, and the qubit's minimum gap frequency $\Delta$ satisfy the relation $[g\gtrsim \max (\omega,\sqrt{\Delta\omega}/2)]$
and the energy eigenstates including the ground state are highly entangled.
This highly entangled ground state is of interest both for its physical novelty,
its implications about the limits of the light-matter interaction strength,
and potentially being used as a quantum resource, e.g. as a robust source of entangled pairs.
Ultrastrong coupling could also be used to implement ultrafast quantum gates~\cite{Ballester12}.
It should be noted that the term deep strong coupling was introduced recently~\cite{Casanova},
before which this regime was treated as part of the ultrastrong-coupling regime.
The experimental progress towards stronger coupling has been accompanied by numerous theoretical studies on the behavior of circuit-QED systems in this regime
\cite{Devoret,Irish,Hausinger,Zueco,Bourassa,Ashhab2010,Hwang2010,Casanova,DelValle,Nataf2010,Viehmann,Beaudoin,
Braak,Ridolfo,Ashhab2013,Law,DeLiberato,Bamba,Hwang2015,Ying, Kyaw15,Jaako16}.
Related recent experimental studies investigated the coupling of a qubit ensemble to a single cavity~\cite{Kakuyanagi}, the coupling of a single qubit to a continuum of modes in a superconducting transmission line~\cite{FornDiaz2016},
and the coupling of a single qubit to a resonant mode of two superconducting resonators~\cite{Baust16}.

The deep-strong-coupling spectra reported in Ref.~\cite{Yoshihara}
are quite different from those seen in the conventional spectra of previously studied circuits, e.g. those of Ref.~\cite{FornDiaz2010}.
One of the remarkable features is that the $|0\rangle \to |2\rangle$ transition has a dip rather than a peak when the qubit is biased at the symmetry point.
Here, $|0\rangle$ stands for the ground state and $|n\rangle$ with $n\geq 1$ stands for the $n$th excited state of the combined system.
Another reamarkable feature is that the $|0\rangle \to |2\rangle$ and $|1\rangle \to |3\rangle$
transitions disappear at the symmetry point.
This observation raises the questions
of how the different features in the spectra are related to the exact value of the coupling strength $g$
and what is the physical origin of each feature.
In order to address these questions
we investigate circuits with intermediate values of $g$, and we complement this experimental investigation with a systematic theoretical analysis of the spectra for the full range of coupling strength values. We find that there are five coupling regimes with qualitatively different spectral features
each having a different physical origin.
In particular, the intermediate coupling strength circuits on which we report here display their own unique spectral features that are different from both weaker  and stronger coupling circuits. 

\section{Hamiltonian and experimental results}
\label{Sec2}

The circuit that have used for this work can be described as a composite system that comprises one flux qubit inductively coupled to a lumped-element LC oscillator via Josephson junctions. The circuit design is similar to that of Ref.~\cite{Yoshihara}, except for the following two differences: (i) To study the intermediate coupling strength regime, the persistent current of the flux qubit is designed to be somewhat smaller than those investigated in Ref.~\cite{Yoshihara}. (ii) To increase the size of the spectral features around the oscillator’s frequency $\omega$ while keeping the qualitative features of the spectra unchanged, the qubit’s minimum gap $\Delta$ is designed to be around $\omega/4$.

The qubit-oscillator circuit is described by the Hamiltonian
\begin{equation}
\hat{H} = -\frac{\Delta}{2} \hat{\sigma}_x-\frac{\epsilon}{2} \hat{\sigma}_z+\omega \hat{a}^{\dagger}\hat{a}+ g \hat{\sigma}_z \left( \hat{a} + \hat{a}^{\dagger} \right).
\label{Eq:Hamiltonian}
\end{equation}
The parameter $\epsilon$ represents the qubit's bias point measured relative to the symmetry point. The operators $\hat{\sigma}_{x,z}$ are qubit Pauli operators, and $\hat{a}$ and $\hat{a}^{\dagger}$ are the oscillator's lowering and raising operators, respectively. Note that we have set $\hbar=1$ here, and we shall do so throughout this paper. We shall also ignore any higher-order terms in the Hamiltonian, such as the so-called $A^2$ term \cite{Rzazewski}. As discussed in Ref.~\cite{Yoshihara}, the $A^2$ term would simply renormalize the system parameters and would therefore not result in spectra that look qualitatively different from those that we shall present here. Any given experimentally observed spectrum can, however, be used to set an upper bound for the $A^2$ term in the Hamiltonian.

\begin{figure}
\includegraphics{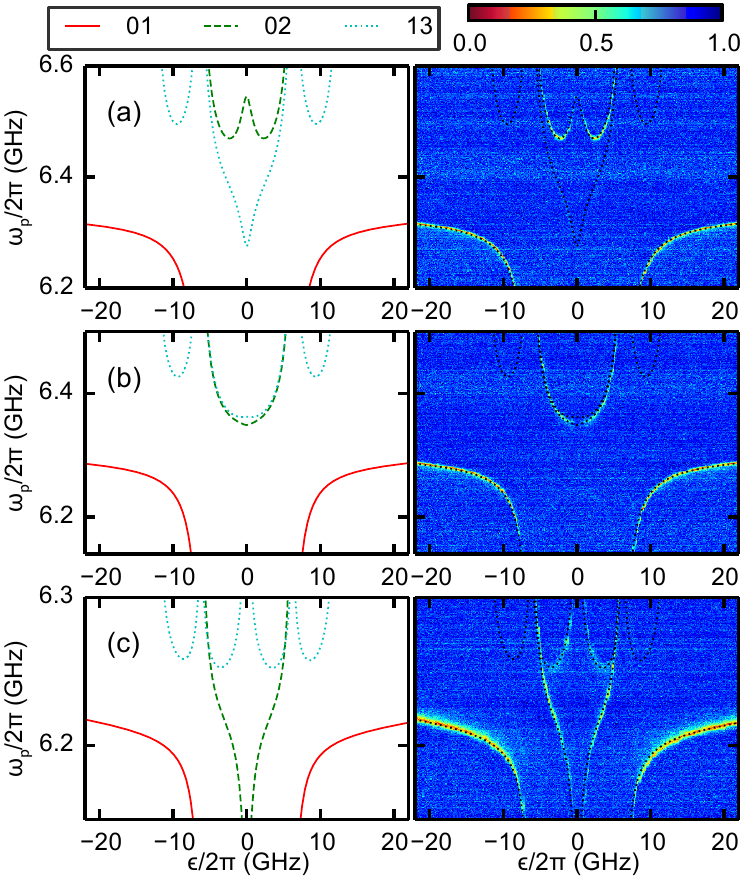}
\caption{Measured transmission spectra for the qubit-oscillator circuit at three different bias points with the parameters: (a) $\Delta/(2\pi)=2.08$ GHz, $\omega/(2\pi)=6.305$ GHz, $g/(2\pi)=4.08$ GHz; (b) $\Delta/(2\pi)=1.85$ GHz, $\omega/(2\pi)=6.275$ GHz, $g/(2\pi)=4.44$ GHz; (c) $\Delta/(2\pi)=1.31$ GHz, $\omega/(2\pi)=6.203$ GHz, $g/(2\pi)=5.31$ GHz. These parameters give $g/\omega=$0.65 (a), 0.71 (b) and 0.86 (c). The left-hand side of each panel shows the central frequencies of the different spectral lines as functions of qubit bias $\epsilon$, calculated from the eigenvalues of the Hamiltonian. The solid red, dashed green and dotted cyan lines correspond to the transitions $\ket{0}\rightarrow\ket{1}$, $\ket{0}\rightarrow\ket{2}$ and $\ket{1}\rightarrow\ket{3}$, respectively.
The right-hand side of each panel shows the measured transmission amplitude as a function of $\epsilon$ and probe frequency $\omega_{\rm p}$.
The color scheme is chosen such that the lowest point in each spectrum is red and the highest point is blue.
The dotted black lines correspond to the same transitions shown in the left-hand side.}
\label{Fig:ExperimentalSpectra}
\end{figure}

Figure \ref{Fig:ExperimentalSpectra} shows the normalized amplitudes of the transmission spectra
$|S_{21}(\epsilon,\omega_{\rm p})|/|S_{21}(\omega_{\rm p})|_{\rm max}$
that we have measured
around three different bias points that all correspond to half-integer values of the qubit loop's flux but,
because of our tunable coupling circuit design~\cite{Yoshihara},
have different values of qubit-oscillator coupling strength (see Appendix A).
Here, $\omega_{\rm p}$ is the probe frequency and $|S_{21}(\omega_{\rm p})|_{\rm max}=\textrm{max}_{\epsilon}|S_{21}(\epsilon,\omega_{\rm p})|$.
The three points correspond to $g/\omega=$~0.65, 0.71 and 0.86. The parameters are obtained from fitting the experimentally measured resonance frequencies to those calculated theoretically by diagonalizing the Hamiltonian given in Eq.~(\ref{Eq:Hamiltonian}) with $\Delta$, $\omega$ and $g$ treated as fitting parameters.

When $g/\omega=0.65$, the spectral line for the $\ket{0}\rightarrow\ket{2}$ transition has a $/\backslash$ shape and the (barely visible) $\ket{1}\rightarrow\ket{3}$ transition has a $\backslash /$ shape around $\epsilon=0$.
When $g/\omega=0.86$, the situation is reversed: the spectral line of the $\ket{0}\rightarrow\ket{2}$ transition has a $\backslash /$ shape and that of the $\ket{1}\rightarrow\ket{3}$ transition has a $/\backslash$ shape around $\epsilon=0$. When $g/\omega=0.71$, there seems to be only one spectral line between $\epsilon=-\omega$ and $\epsilon=\omega$, and this line has a broad
U shape between $\epsilon=-\omega$ and $\epsilon=\omega$ with a small gap at $\epsilon=0$. We shall show below that all of these features are characteristic features of the respective coupling regimes.

\section{Calculated spectra}
\label{Sec3}

In this section we use theoretical and numerical calculations to explain the different features in the measured spectra.
In particular we show that the range $0<g/\omega<\infty$ can be divided into five intervals, namely
[0, 0.38], [0.38, 0.5],
[0.5, 0.71], [0.71, 0.92] and [0.92, $\infty$];
with each one of these intervals having its own characteristic spectrum.
These characteristic spectra can be understood from the dependence of the energy levels on the coupling strength as well as the symmetry of the energy eigenstates.
The experimental spectra shown here, in combination with those of Refs.~\cite{FornDiaz2010,Yoshihara}, cover four of these five intervals and in all cases the features that can be resolved by simple visual inspection of the spectra are consistent with the exact parameter values extracted from a systematic fitting of the full spectra. Unfortunately, none of our circuits turned out to have
$0.38<g/\omega<0.5$,
and therefore we do not have an experimental example of the corresponding pattern.
 
For the calculation of the spectra, we start by considering the transmission and reflection
of a two-level quantum system that is kept at a sufficiently low temperature and probed sufficiently weakly such that it can be assumed to be in its ground state most of the time.
The dynamics
and steady state of such a system can be described using the optical
Bloch equations~\cite{CohenTan}.
The system partially reflects
an external driving probe field, and the reflection coefficient is
described by the formula
\begin{equation}
R(\omega_{\rm p}) =R_0 \frac{\Omega_{01}^2}{\Omega_{01}^2+\left(\omega_{\rm p}-\omega_{01}\right)^2 + \Gamma^2},
\label{Eq:ResponseOfSingleQubit}
\end{equation}
where
\begin{eqnarray}
\Omega_{01} & = & \left| A_{\rm p} \times \bra{1}\hat{x}\ket{0} \right|, \nonumber \\
\omega_{01} & = & E_1-E_0,
\label{Eq:OmegaDefinitions}
\end{eqnarray}
$R_0$ sets the maximum value of the reflection coefficient
at $\omega_{\rm p} = \omega_{01}$,
$A_{\rm p}$ is the amplitude of the driving probe field, $\hat{x}$ is the system operator that is driven by the probe field, which in our experiment is $\left( \hat{a} + \hat{a}^{\dagger} \right)$
(such that the interaction with the probe field can be described by the Hamiltonian $\hat{H}_{\rm p}=A_{\rm p} \hat{x}\cos\left[\omega_{\rm p}t\right]/2$), and $E_n$ is the energy of the state $\ket{n}$.
The transmission coefficient is given by $T = 1-R$.
The parameter $\Gamma$ represents an overall decoherence rate in the system.
The above formula can be seen as resulting from a simplified description of the interplay between multiple physical processes, and more complex expressions could be derived using alternative derivations
(e.g. separating the effects of zero-, low- and high-frequency noise into
different decoherence channels).
However, this formula gives all of the overall features of the
transmission spectra
that interest us in this work, and we therefore do not complicate the picture with any additional elements.
The main role of the decoherence parameter $\Gamma$ in our
calculations is to set width of the spectral lines, assuming a small
value of $\Omega_{01}$.

Generalizing Eq.~(\ref{Eq:ResponseOfSingleQubit}) to a multi-level quantum system, we obtain for the reflection coefficient
\begin{equation}
R(\omega_{\rm p}) =R_0 \sum_{i,j (j>i)} P_i \frac{\Omega_{ij}^2}{\Omega_{ij}^2+\left(\omega_{\rm p}-\omega_{ij}\right)^2 + \Gamma^2},
\label{Eq:MultiLevelResponse}
\end{equation}
where the indices $i$ and $j$ run over all the quantum states of the system (with $j>i$), $P_i$ is the thermal-equilibrium occupation probability of state $\ket{i}$, and the definitions of $\Omega_{ij}$ and $\omega_{ij}$ follow straightforwardly from Eq.~(\ref{Eq:OmegaDefinitions}). Note that in general $\Gamma$ will not be the same for all the transitions, but calculating accurate values for the different transitions will not affect the main phenomena that we wish to study here, and we therefore ignore this point for simplicity.

We now present theoretically calculated spectra that illustrate the different transformations in the spectral features.
As we shall explain below,
the features can be understood from the dependence of the energy levels on the coupling strength,
in particular crossings between energy levels and other crossings between transition frequencies,
as well as the symmetry of the energy eigenstates.
Some of our calculations were performed using the QuTiP simulation package~\cite{QuTiP}.

For definiteness we set $\Delta/\omega=0.1$, which is sufficiently small that various results can be understood in terms of our knowledge about the small $\Delta/\omega$ limit without losing any of the relevant features in the spectra. For the calculation of the occupation probabilities $P_i$, we use the temperature setting $k_{\rm B}T/\omega=0.5$, which is a relatively high temperature compared to what can be achieved in state-of-the-art experiments, but it makes certain spectral features more easily visible. With such temperatures, the states $\ket{0}$ and $\ket{1}$ are generally both significantly populated in the bias range of interest, while the occupation probabilities of higher levels are negligible. This situation is consistent with the majority of the data sets reported here and in Ref.~\cite{Yoshihara}. We set $A_{\rm p}=2\times10^{-3}$ and $\Gamma=3\times 10^{-3}$. These parameters set the amplitudes and widths of the spectral lines. We focus on the frequency range around $\omega_{\rm p}=\omega$, because this range gives the strongest spectroscopic response, although in principle one could probe the system at other frequencies in order to obtain more information about the system parameters.

\begin{figure*}
\includegraphics{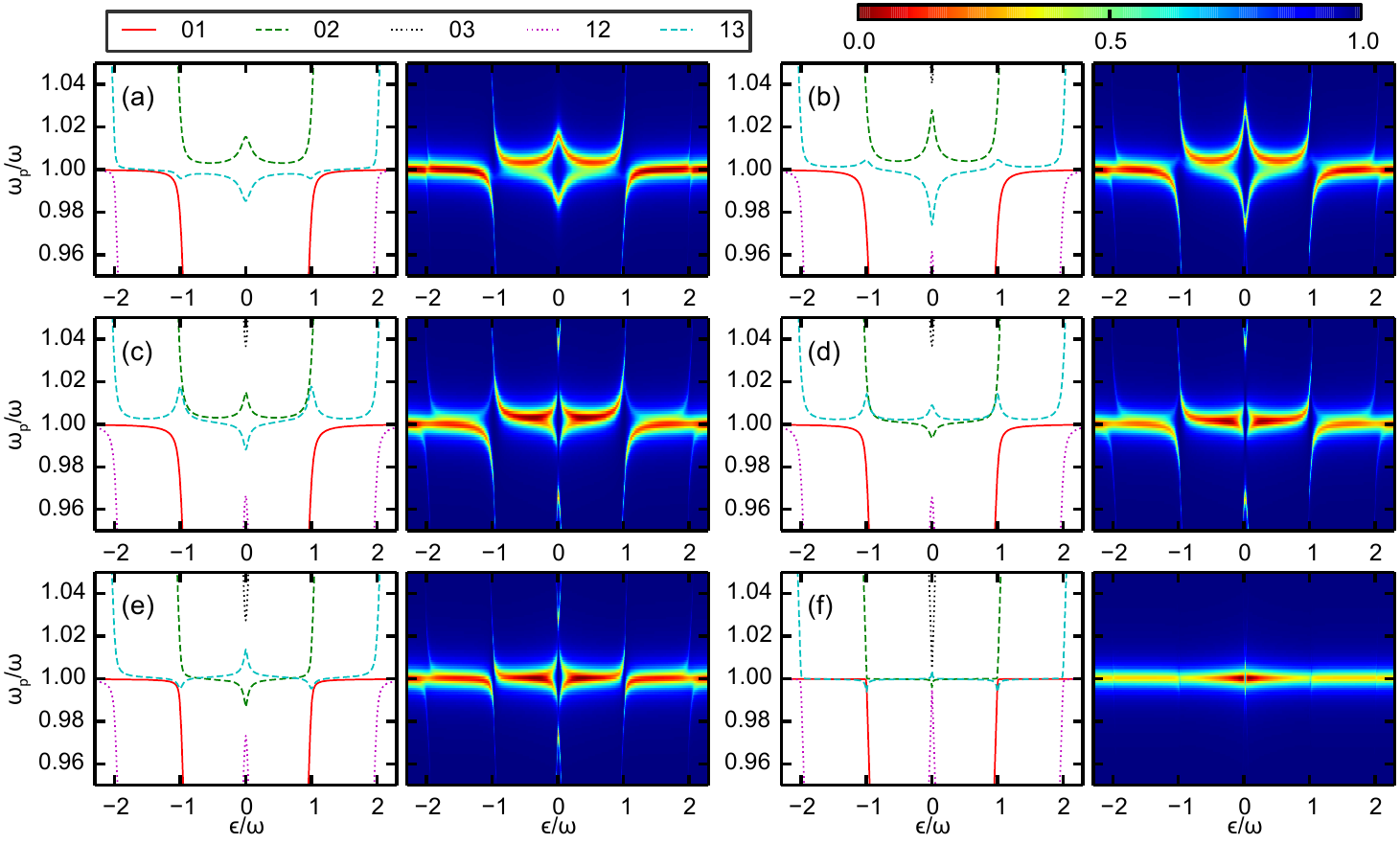}
\caption{Calculated transmission spectra with increasing coupling strength for $\Delta/\omega=0.1$. The different panels correspond to $g/\omega=$0.3 (a), 0.45 (b), 0.6 (c), 0.8 (d), 1.0 (e) and 1.5 (f). The left-hand side of each panel shows the central frequencies of the different spectral lines as functions of qubit bias $\epsilon$, calculated from the eigenvalues of the Hamiltonian. The solid red, dashed green, dotted black, dotted cyan and dashed magenta lines correspond to the transitions $\ket{0}\rightarrow\ket{1}$, $\ket{0}\rightarrow\ket{2}$, $\ket{0}\rightarrow\ket{3}$, $\ket{1}\rightarrow\ket{2}$ and $\ket{1}\rightarrow\ket{3}$, respectively.
The right-hand side of each panel shows the
transmission coefficient $1-R$ (with $R$ calculated using Eq.~\ref{Eq:MultiLevelResponse}, where we assume $R_0 = 1$)
as a function of $\epsilon$ and probe frequency $\omega_{\rm p}$.
}
\label{Fig:SpectraFrom01}
\end{figure*}

Figure \ref{Fig:SpectraFrom01} shows the theoretically calculated spectra for six different values of the coupling strength $g$ that are representative of the qualitatively different regimes that can be obtained with this system. To directly compare the calculated spectra with the experimentally measured transmission spectra,
we plot $1-R$ assuming $R_0 = 1$.

When $g/\omega = 0.3$
[Fig.~\ref{Fig:SpectraFrom01}(a)], the spectral line of the $\ket{0}\rightarrow\ket{2}$ transition has a W shape in the range $-\omega<\epsilon<\omega$, while that of the $\ket{1}\rightarrow\ket{3}$ transition has an M shape. The only qualitative change that takes place in going from $g/\omega=0$ to $g/\omega \sim 0.3$ is that the distances of these spectral lines from the central frequency $\omega_{\rm p}=\omega$ increase with increasing $g/\omega$. The transition from the weak-coupling regime to the strong-coupling regime of circuit-QED usually occurs in this interval.
Furthermore, as mentioned in Sec.~I, recently the term ultrastrong coupling has come to be used for the regime with $g\sim 0.1\omega$, such that the ultrastrong-coupling regime also exhibits this spectral pattern. Indeed, the spectrum observed in Ref.~\cite{FornDiaz2010} resembles that shown in Fig.~\ref{Fig:SpectraFrom01}(a), although in that experiment the signal from the $\ket{1}\rightarrow\ket{3}$ transition was too weak and its exact $\epsilon$ dependence cannot be inferred from the experimental data.

When $g/\omega$ changes from 0.3 to 0.45 [Figs.~\ref{Fig:SpectraFrom01}(a) and (b)] a more serious change occurs in the sense that the change can be seen plainly without an examination of the exact values of the different frequencies. The sides of the M-shaped $\ket{1}\rightarrow\ket{3}$ spectral line go up and the line transforms into a V shape, which is most clearly seen in the plots of the central frequencies
[i.e. the left-hand sides of Figs.~\ref{Fig:SpectraFrom01}(a) and (b)].
This qualitative change in the shape of the spectral line
occurs at $g/\omega \sim 0.383$ and
can therefore be used as an indicator of whether the ratio $g/\omega$ is smaller or larger than 0.383. 
This transformation in the spectrum can be understood as follows: at
the bias point $\epsilon=\omega$ (and assuming a small value of
$\Delta$), the energy level ladder has a nondegenerate ground state $|0\rangle$
followed by pairs of nearly degenerate of energy levels. The
pair $\{\ket{1},\ket{2}\}$ have energies given by \cite{Ashhab2010}
\begin{eqnarray}
E_{1,2} - E_0 & = & \omega \pm e^{-2g^2/\omega^2} \frac{2g}{\omega}
L_0^1\left[\frac{4g^2}{\omega^2}\right] \nonumber \\
& = & \omega \pm e^{-2g^2/\omega^2} \frac{2g}{\omega},
\label{Eq:E12-E0}
\end{eqnarray}
while the pair $\{\ket{3},\ket{4}\}$ have energies given by
\begin{eqnarray}
E_{3,4} - E_0 & = & 2 \omega \pm e^{-2g^2/\omega^2} \frac{2g}{\omega}
\frac{1}{\sqrt{2}} L_1^1\left[\frac{4g^2}{\omega^2}\right] \nonumber \\
& = & 2 \omega \pm e^{-2g^2/\omega^2} \frac{2g}{\omega}
 \frac{1}{\sqrt{2}} \left(2-\frac{4g^2}{\omega^2}\right).
\label{Eq:E34-E0}
 \end{eqnarray}
Here $L_n^m(x)$ are associated Laguerre polynomials, and we have used
the facts that $L_0^1(x)=1$ and $L_1^1(x)=2-x$. In
Eq.~(\ref{Eq:E12-E0}) the plus sign gives $E_2-E_0$ while the minus sign
gives $E_1-E_0$. In Eq.~(\ref{Eq:E34-E0}), for $g/\omega<1/\sqrt{2}$ the
plus sign gives $E_4-E_0$ while the minus sign gives $E_3-E_0$, and the sign
assignment is reversed for $g/\omega>1/\sqrt{2}$. Combining
Eqs.~(\ref{Eq:E12-E0}) and (\ref{Eq:E34-E0}), we find that the
frequency of the transition $\ket{1}\rightarrow\ket{3}$ is given by
 \begin{equation}
 E_{3} - E_1 = \omega + e^{-2g^2/\omega^2} \frac{2g}{\omega} \left[ 1
 \pm \frac{1}{\sqrt{2}} \left(2-\frac{4g^2}{\omega^2}\right) \right].
 \label{Eq:E3-E1}
 \end{equation}
 For $g/\omega<1/\sqrt{2}$ the minus sign should be used. The second
 term in this formula changes sign from negative to positive at
 $g/\omega=\sqrt{2-\sqrt{2}}/2=0.383$,
which explains why the dip in the spectrum turns into a peak at this value of $g/\omega$.

Another feature in the spectrum that can exhibit different properties depending on the coupling strength is the continuity of the spectral lines at the symmetry point $\epsilon=0$.
From $g/\omega=0$ up to $g/\omega=0.5$, the $\ket{0}\rightarrow\ket{2}$ and $\ket{1}\rightarrow\ket{3}$ spectral lines are continuous across the point $\epsilon=0$, indicating that the transitions are allowed at the symmetry point.
In the same interval, both of these lines move away from the central frequency $\omega_{\rm p}=\omega$ with increasing $g/\omega$. Meanwhile, the spectral line of the $\ket{0}\rightarrow\ket{3}$ transition, which has a V shape, approaches the $\ket{0}\rightarrow\ket{2}$ line from above, and the $\ket{0}\rightarrow\ket{3}$ line vanishes at the symmetry point, indicating that it corresponds to a forbidden transition at the symmetry point.
The transition is forbidden because the states $\ket{0}$ and $\ket{3}$
have different symmetries. The states are given by
\begin{eqnarray}
\ket{0} & = & \frac{1}{\sqrt{2}} \left( \ket{L}_{\rm q} \otimes
\hat{D}(-\alpha)\ket{0}_{\rm o} +  \ket{R}_{\rm q} \otimes
\hat{D}(\alpha)\ket{0}_{\rm o} \right) \nonumber \\
\ket{3} & = & \frac{1}{\sqrt{2}} \left( \ket{L}_{\rm q} \otimes
\hat{D}(-\alpha)\ket{1}_{\rm o} - \ket{R}_{\rm q} \otimes \hat{D}(\alpha)\ket{1}_{\rm o}
\right),
\label{Eq:Eigenstates03}
\end{eqnarray}
where the states $\ket{L}_{\rm q}$ and $\ket{R}_{\rm q}$ are the eigenstates of
$\hat{\sigma}_z$, the states $\ket{0}_{\rm o}$ and $\ket{1}_{\rm o}$ are,
respectively, the oscillator states containing 0 and 1 photons, and
$\hat{D}(\alpha)=\exp (\alpha\hat{a}^{\dagger}-\alpha^*\hat{a})$.
Because the first and second lines in Eq.~(\ref{Eq:Eigenstates03})
have opposite signs between the two terms of each one, the matrix
element
$\bra{0}(\hat{a}+\hat{a}^{\dagger})\ket{3}$ vanishes and the
transition $\ket{0}\rightarrow\ket{3}$ is forbidden.
Similarly, the spectral line of the $\ket{1}\rightarrow\ket{2}$ transition, which has a $\Lambda$ shape, approaches the $\ket{1}\rightarrow\ket{3}$ line from below, and it vanishes at the symmetry point. At $g/\omega=0.5$, which is the point where $E_2=E_3$,
the states $\ket{2}$ and $\ket{3}$ undergo an energy level crossing
such that they swap their physical properties.
At the point of the energy level crossing, the frequencies of the two transitions in each pair described above coincide with each other,
and the corresponding spectral lines touch each other.
Above $g/\omega=0.5$, the pairs of lines separate again and the inner two lines start to move back towards the central frequency $\omega_{\rm p}=\omega$. Importantly, above $g/\omega=0.5$ the $\ket{0}\rightarrow\ket{2}$ and $\ket{1}\rightarrow\ket{3}$ transitions are forbidden at the symmetry point while the $\ket{0}\rightarrow\ket{3}$ and $\ket{1}\rightarrow\ket{2}$ transitions are allowed. This property can be seen plainly in
the right-hand side of each panel of
Figs.~\ref{Fig:SpectraFrom01}(c)--(f) by observing that the $\ket{0}\rightarrow\ket{2}$ and $\ket{1}\rightarrow\ket{3}$ lines are now discontinuous at $\epsilon=0$. All the
experimental spectra presented here and in Ref.~\cite{Yoshihara} exhibit this feature, and we can therefore say that in all of these cases $g/\omega>0.5$.

As we increase $g/\omega$ above 0.5, the peak in the $\ket{0}\rightarrow\ket{2}$ line at $\epsilon=0$ continues to go down and the dip in the $\ket{1}\rightarrow\ket{3}$ line continues to go up until the two lines cross each other at $g/\omega=1/\sqrt{2}\approx0.71$.
This point corresponds to the condition $E_3-E_2=E_1-E_0$, which explains the degeneracy in the frequencies of the
$\ket{0}\rightarrow\ket{2}$ and $\ket{1}\rightarrow\ket{3}$
spectral lines.
The spectrum plotted in Fig.~\ref{Fig:ExperimentalSpectra}(b) has all the features expected for a coupling strength that is close to this point, and indeed the parameters that we obtain from fitting the full spectrum give $g/\omega=0.71$, which is very close to the transformation point.
Note that there are no energy level crossings or avoided crossings at
this point. It is just the energy level differences that cross each
other.
When $g/\omega>0.71$ [e.g.~as in Fig.~\ref{Fig:SpectraFrom01}(d)], the $\ket{0}\rightarrow\ket{2}$ line has a V shape, while the $\ket{1}\rightarrow\ket{3}$ line has a W shape in the range $-\omega<\epsilon<\omega$.
The spectrum in Fig.~\ref{Fig:ExperimentalSpectra}(c) displays this pattern and we can therefore say that $g/\omega>0.71$ in this case, in agreement with what we obtain from fitting the full spectrum.

If we increase $g/\omega$ further, we find that at
$g/\omega=0.924$
the two edges of the W-shaped $\ket{1}\rightarrow\ket{3}$ line go down and the line transforms into a $\Lambda$ shape.
This transformation can be understood from Eq.~(\ref{Eq:E3-E1}) with the plus sign.
Now the second term changes sign from positive to
negative at $g/\omega=\sqrt{2+\sqrt{2}}/2=0.924$, and hence the
$\ket{1}\rightarrow\ket{3}$ transition line exhibits a transformation
from having a peak to having a dip at $\epsilon=\omega$.
Note that although the peaks and dips in Figs.~\ref{Fig:SpectraFrom01}(d) and (e) look small,
suggesting that they might be difficult to observe experimentally,
this appearance is a result of the fact that we use the same range for the y axis in all the panels.
The spectrum in Fig.~2\textbf{d} of Ref.~\cite{Yoshihara} clearly exhibits
a dip in the $\ket{1}\rightarrow\ket{3}$ spectral line as $\epsilon$ approaches $\omega$ as would be expected for its very strong coupling ($g/\omega=1.34$).
As the coupling strength is increased in the regime
$g/\omega> 0.92$,
no further qualitative changes occur in the spectrum, except for the fact that the features decrease in size as $g/\omega$ increases, such that eventually for very strong coupling we recover only an $\epsilon$-independent spectral response at the bare oscillator frequency, i.e.~at $\omega_{\rm p}=\omega$.
In Fig.~\ref{Fig:SpectraFrom01}(f), the $|0\rangle \to |2\rangle$, $|0\rangle \to |3\rangle$,
$|1\rangle \to |2\rangle$, and $|1\rangle \to |3\rangle$ spectral lines all collapse to $\omega_{\rm p}/\omega = 1$ at $\epsilon = 0$,
and the discontinuity of the $|0\rangle \to |2\rangle$ and $|1\rangle \to |3\rangle$ spectral lines at $\epsilon = 0$, where these transitions are forbidden, is covered by the $|0\rangle \to |2\rangle$ and $|1\rangle \to |3\rangle$ spectral lines, which are both allowed at $\epsilon = 0$.
The reason behind this behavior is that in the limit of large $g/\omega$,
the qubit-oscillator correlations in the low-lying states become extremely strong,
such that no effect of the superpositions involving different values of $\hat{\sigma}_z$ can be observed.
In other words, the first term in the Hamiltonian (proportional to
$\hat{\sigma}_x$) can be ignored. In this case, regardless of the
value of $\hat{\sigma}_z$, the system behaves as a qubit in a fixed state
imparting a constant force on a harmonic oscillator, which does not
affect the spectral response of the oscillator.

The above discussion leads to a quick and simple method for obtaining a rough estimate of the parameter $g/\omega$ and identifying in which of the five intervals it lies simply by looking at the overall features in the spectrum. We also note that since the central frequency in the spectrum gives the oscillator's bare frequency $\omega$, the estimate that we obtain for the ratio $g/\omega$ immediately gives us an estimate for $g$.

\section{Conclusion}

We have measured spectra of a circuit-QED system that exhibits unique features different from those observed in previous experiments. We have also performed a systematic analysis of the expected spectra for different values of the coupling strength and demonstrated that there are several possible spectral patterns depending on the coupling strength. In other words, the features in the spectrum undergo several qualitative transformations as the qubit-oscillator coupling strength is increased. The various features can be used to identify the coupling regime and estimate certain parameters with reasonable accuracy, even without any quantitative fitting of experimental data. Furthermore, it could happen that technical issues such as high dissipation rates could broaden spectral lines to the extent that fitting the data becomes unreliable for extracting the values of multiple parameters. In such cases, one could nevertheless use the qualitative method that we have presented here to estimate the parameters based on the shapes of a few overall features in the spectrum. Our results are a further demonstration of the richness of the physics that results from the simple circuit-QED Hamiltonian and can find applications in future experiments that push the parameters of such systems further into yet-unexplored regimes.

\begin{acknowledgements}
We thank Masao Hirokawa, Kae Nemoto, John W. Munro, Yuichiro Matsuzaki, and Motoaki Bamba for stimulating discussions.
This work was supported in part by the Scientific Research (S) Grant No.~JP25220601 by the Japan Society for the Promotion of Science (JSPS). 
\end{acknowledgements}

\appendix
\section{experimental setup}
\begin{figure}
\includegraphics{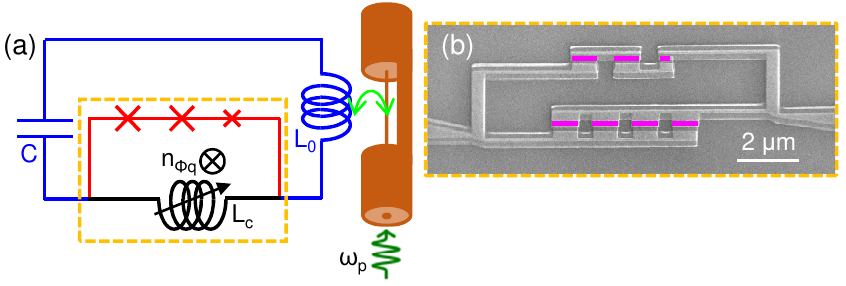}
\caption{(a) Circuit diagram.
A superconducting flux qubit (red and black) and a superconducting LC oscillator (blue and black)
are inductively coupled to each other by sharing a tunable inductance (black).
(b) Scanning electron microscope image of the qubit including the coupler junctions located at the orange rectangle in diagram (a).
Josephson junctions are indicated as magenta rectangles.
The coupler, consisting of four parallel Josephson junctions,
is tunable via the magnetic flux bias through its loop.}
\label{Fig:SEMCircuit}
\end{figure}

Figure~\ref{Fig:SEMCircuit}(a) shows a diagram of the qubit-oscillator circuit.
The flux bias through the qubit loop, $n_{\phi}$ is normalized in units of the superconducting flux quantum,
$\Phi_0 = h/2e$.
The energy bias between two persistent current states of a flux qubit is given as
$\epsilon = 2I_{\rm p}\Phi_0(n_{\phi}-n_{\phi 0})$,
where $I_{\rm p}$ is the maximum persistent current,
$n_{\phi 0} = 0.5 + k_{\rm q}$,
and $k_{\rm q}$ is the integer that minimizes $|n_{\phi}-n_{\phi 0}|$.
One may think that higher energy levels of the flux qubit might contribute the energy spectra of the qubit-oscillator circuit as discussed in the Ref.~\cite{Yamamoto2014},
especially in the case of very strong coupling to the oscillator.
The energy of the second excited state in the typical flux qubit is more than 20~GHz,
well larger than the energies of the qubit and oscillator.
Together with the fact that the energy spectra involving up to the third excited states of the qubit-oscillator circuit are well fitted by the Hamiltonian in Eq.~\ref{Eq:Hamiltonian},
the contribution from the second or higher excited states might modify the sample parameters,
but does not change the shape of the Hamiltonian.
Note that the bare sample parameters in the case of zero coupling cannot be experimentally determined,
and the amount of the modifications of the sample parameters cannot be evaluated.

Figure~\ref{Fig:SEMCircuit}(b) shows a scanning electron microscope image of the qubit
including the coupler.
The coupler consists of four parallel Josephson junctions.
The critical current of the coupler is given as
\begin{equation}
I_{\rm c(coup)} = 4I_{\rm c}\cos (2\pi n_{\phi \rm c})\cos (\pi n_{\phi \rm c}),
\label{Eq:Ic_coupler}
\end{equation}
where $I_{\rm c}$ is the critical current of each Josephson junction,
and $n_{\phi \rm c}$ is the normalized flux bias through each coupler loop defined by two neighboring parallel junctions.
An external superconducting magnet produces uniform magnetic field,
and magnetic fluxes are applied to the qubit and the coupler proportional to the area of their loops.
The area ratio of the loops $r_{\rm c} = A_{\rm coupler}/A_{\rm qubit}$ is approximately 0.05,
where $A_{\rm coupler}$ and $A_{\rm qubit}$ are the loop areas of the qubit and the coupler, respectively.
The flux bias through each coupler loop $n_{\phi \rm c} = r_{\rm c}n_{\phi}$ depends on $n_{\phi}$,
which in most cases is around the symmetric point, i.e. $n_{\phi} = \pm 0.5, \pm 1.5, \pm 2.5$ and so on.

Spectroscopy was performed by measuring the transmission spectrum through a coplanar transmission line that is inductively coupled to the LC oscillator.
When the frequency of the probe signal $\omega_{\rm p}$ matches the frequency of a transition between two energy levels,
the transmission amplitude decreases, provided that the transition matrix element is not zero.
The line width of the $|0\rangle \to |3\rangle$ transition at $\epsilon = 0$ for $g/\omega = 0.86$ (not shown) is 6.9~MHz $\times 2\pi$, which can be related to $\Gamma$.
The maximum value of the reflection coefficient $R_0$ depends on the quality factor ratio $Q/Q_{\rm e}$,
where $Q$ is the total quality factor of the qubit-oscillator circuit,
and $Q_{\rm e}$ is the external quality factor of the qubit-oscillator circuit to the coplanar transmission line~\cite{Khalil2012JAP}.
The samples are measured in a dilution refrigerator with a nominal base temperature of 20~mK.

\section{higher gap qubit}
\begin{figure*}
\includegraphics{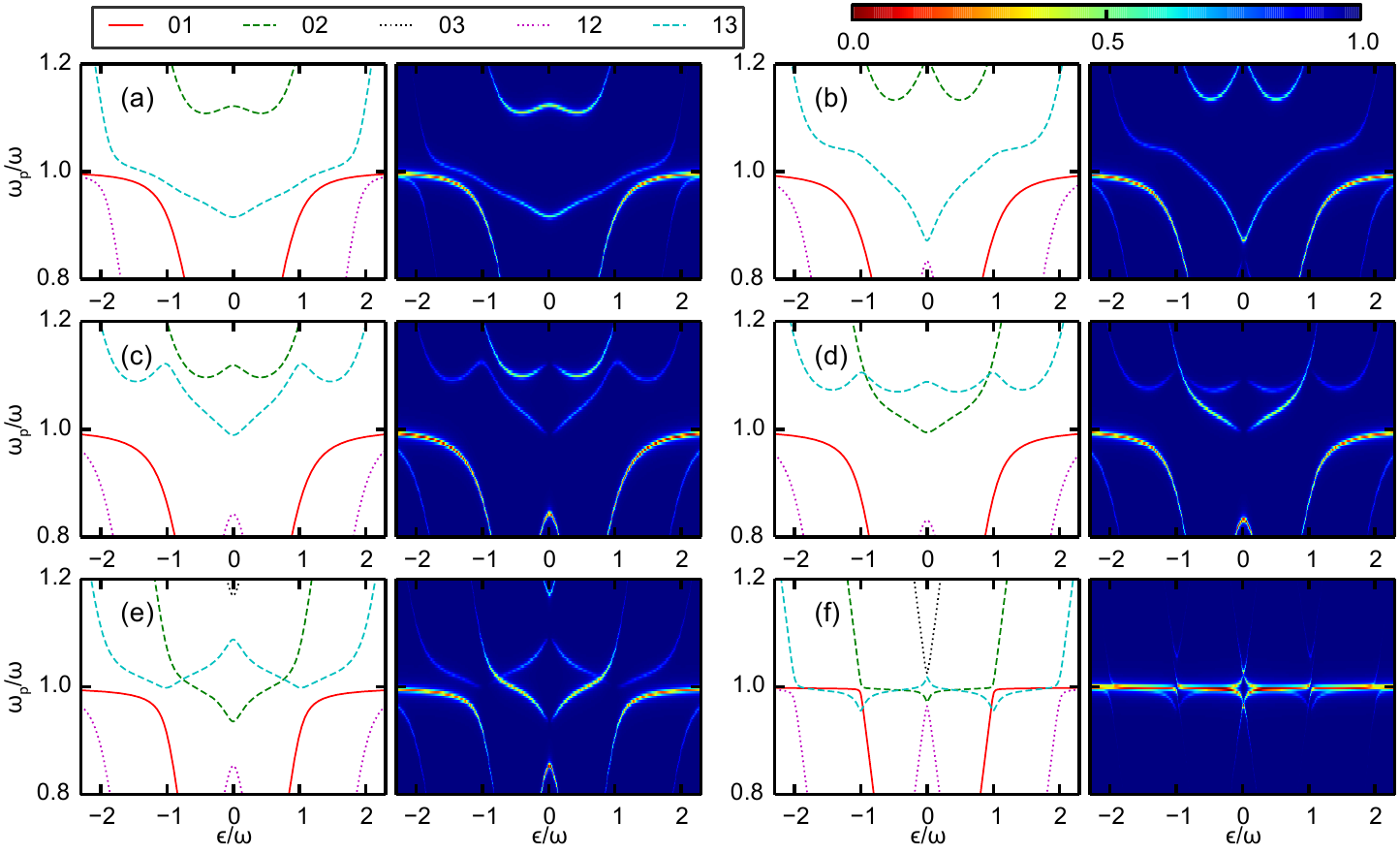}
\caption{Calculated transmission spectra with increasing coupling strength for $\Delta/\omega=0.6$. As in Fig.~2 in the main text, the different panels correspond to $g/\omega=$0.3 (a), 0.45 (b), 0.6 (c), 0.8 (d), 1.0 (e) and 1.5 (f).
The color scheme is chosen such that the lowest point in each spectrum is red and the highest point is blue.}
\label{Fig:SpectraIntermediateDelta}
\end{figure*}

In Fig.~\ref{Fig:SpectraIntermediateDelta} we plot spectra similar to those plotted in Fig.~2 in the main text for a somewhat larger value of $\Delta$, namely $\Delta/\omega=0.6$, which is comparable to the value in circuit III of Ref.~\cite{Yoshihara}. The general description and shape transformations that we have described for the case of small $\Delta$ are also observed in this case, but the fact that the peaks and dips around the symmetry point become larger distorts some of the features discussed in the main text. In particular, we no longer have the M-shaped $\ket{1}\rightarrow\ket{3}$ line at $g/\omega=0.3$, making the transformation around $g/\omega=0.4$ difficult to identify as easily as in the case of small $\Delta$. All of the other shape transformations discussed in Secs.~\ref{Sec2} and \ref{Sec3} are still visible with this value of $\Delta/\omega$. In particular, the spectrum in Fig.~\ref{Fig:SpectraIntermediateDelta}(e) (where $g/\omega=1$) resembles that in circuit III of Ref.~\cite{Yoshihara}, which had $g/\omega=1.01$. Hence we could have obtained a good estimate for $g/\omega$ in that circuit by simply comparing the measured spectrum with the different spectra plotted in the different panels of Fig.~\ref{Fig:SpectraIntermediateDelta}. Rather surprisingly, the pattern transformations that occurred at 0.5 and 0.71 in the small-$\Delta$ case still occur at almost the same points, being shifted to 0.477 and 0.694, respectively.

\begin{figure*}
\includegraphics{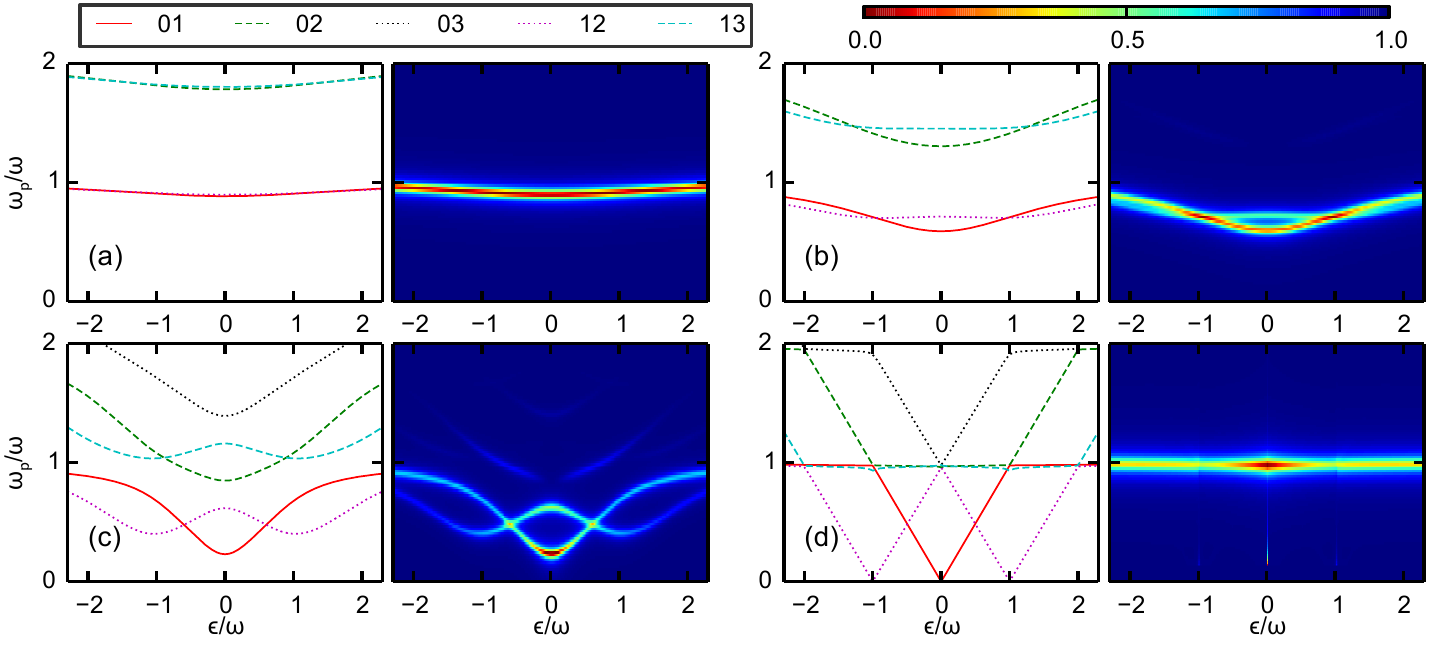}
\caption{Calculated transmission spectra with increasing coupling strength for $\Delta/\omega=3$. The different panels correspond to $g/\omega=$0.4 (a), 0.8 (b), 1.2 (c) and 2 (d). Because the frequency range that we cover in this case is larger than those in the other two cases, here we set $A_{\rm p}=0.02$ and $\Gamma=0.03$. The color scheme is chosen such that the lowest point in each spectrum is red and the highest point is blue.
In panel (d) we have removed an unphysical large peak at $\omega_{\rm p} = \epsilon = 0$.}
\label{Fig:SpectraLargeDelta}
\end{figure*}

In Fig.~\ref{Fig:SpectraLargeDelta} we plot a few different spectra for the case of large $\Delta$, specifically for $\Delta/\omega=3$. The avoided crossings at $\epsilon=\pm\omega$ that one can expect when $\Delta<\omega$ are clearly lost, and the spectrum has fewer shape transformations compared to the two other cases discussed above.
Nevertheless, there are a few quantities that one
can easily read off from the spectrum, such as $\omega_{01}$ far away
from the symmetry point (which gives $\omega_{\rm o}$), $\omega_{01}$
and $\omega_{12}$ at the symmetry point and the value of $\epsilon$ at
which $\omega_{01}$ and $\omega_{12}$ cross. One can then use these
quantities to obtain the values of $g$ and $\Delta$. In general there
are no analytic expressions that can be used to calculate $g$ and
$\Delta$ from the above-mentioned quantities, but a numerical
calculation should be straightforward.

\section{higher energy levels}

\begin{figure}
\includegraphics{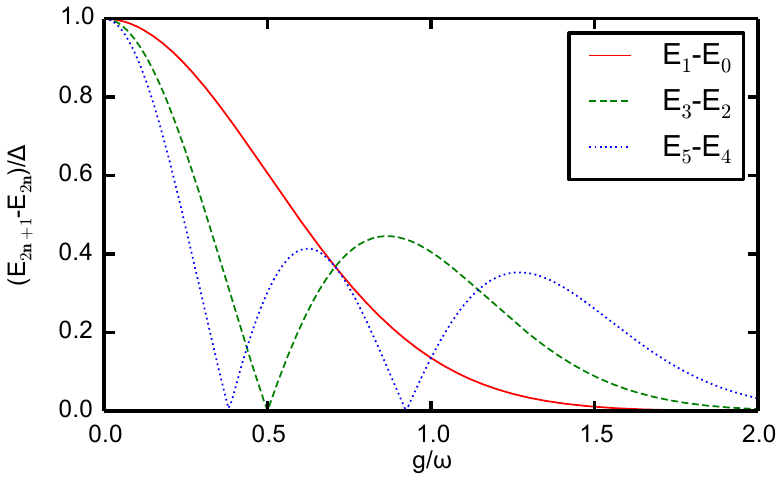}
\caption{Energy level separations $(E_{2n+1}-E_{2n})/\Delta$ with $n=0$ (solid red line), 1 (dashed green line) and 2 (dotted blue line) at $\epsilon=0$ as functions of the ratio $g/\omega$ for $\Delta/\omega=0.1$. Zeros in $E_3-E_2$ and crossings between $E_3-E_2$ and $E_1-E_0$ mark qualitative transformations in the shape of the spectrum as discussed in Secs.~\ref{Sec2} and \ref{Sec3}. The larger number of zeros in $E_5-E_4$ and the larger number of crossings between $E_5-E_4$ and $E_3-E_2$ suggest that spectral lines involving higher energy levels will exhibit a larger number of shape transformations as the coupling strength is varied.}
\label{Fig:EnergyLevelSeparation}
\end{figure}

We have shown that the shapes of the spectral lines corresponding to the $\ket{0}\rightarrow\ket{2}$ and $\ket{1}\rightarrow\ket{3}$ transitions can be used to give a rough estimate for the coupling strength $g$.
In this section we show that the spectral lines corresponding to transitions among higher energy levels can give a more accurate estimate.
In practice, these higher levels could be populated
by artificially heating up the system to higher temperatures.
The thermal energy should be comparable to the photon energy in the oscillator $\hbar \omega$.
In the case $\omega/2\pi = 6.0$~GHz,
the corresponding temperature is given as $T = \hbar \omega/k_{\rm B}$ = 290~mK.
Perhaps a more realistic way to populate the higher levels is to pump
transitions from the lowest energy levels to certain targeted energy levels that then serve as the initial states for transitions to even higher energy levels.

The idea behind the possibility of gaining extra accuracy from utilizing higher levels can be understood based on Fig.~\ref{Fig:EnergyLevelSeparation}, where we plot the energy differences $E_1-E_0$, $E_3-E_2$ and $E_5-E_4$ at $\epsilon=0$ as functions of $g/\omega$. By inspecting $E_1-E_0$ and $E_3-E_2$, we can see that two out of the four spectrum transformations that we discussed in Secs.~\ref{Sec2} and \ref{Sec3} (specifically the shape transformations that occur at $\epsilon=0$) correspond to either a zero in $E_3-E_2$ or a crossing of $E_1-E_0$ and $E_3-E_2$ at $\epsilon=0$. The two other transformations occur at $\epsilon=\pm\omega$ and therefore do not correspond to any special feature in Fig.~\ref{Fig:EnergyLevelSeparation} (in fact those two features seem to coincide with zeros of $E_5-E_4$ at $\epsilon=0$, but we suspect that this is a coincidence that does not have a deep physical origin). 

\begin{figure*}
\includegraphics{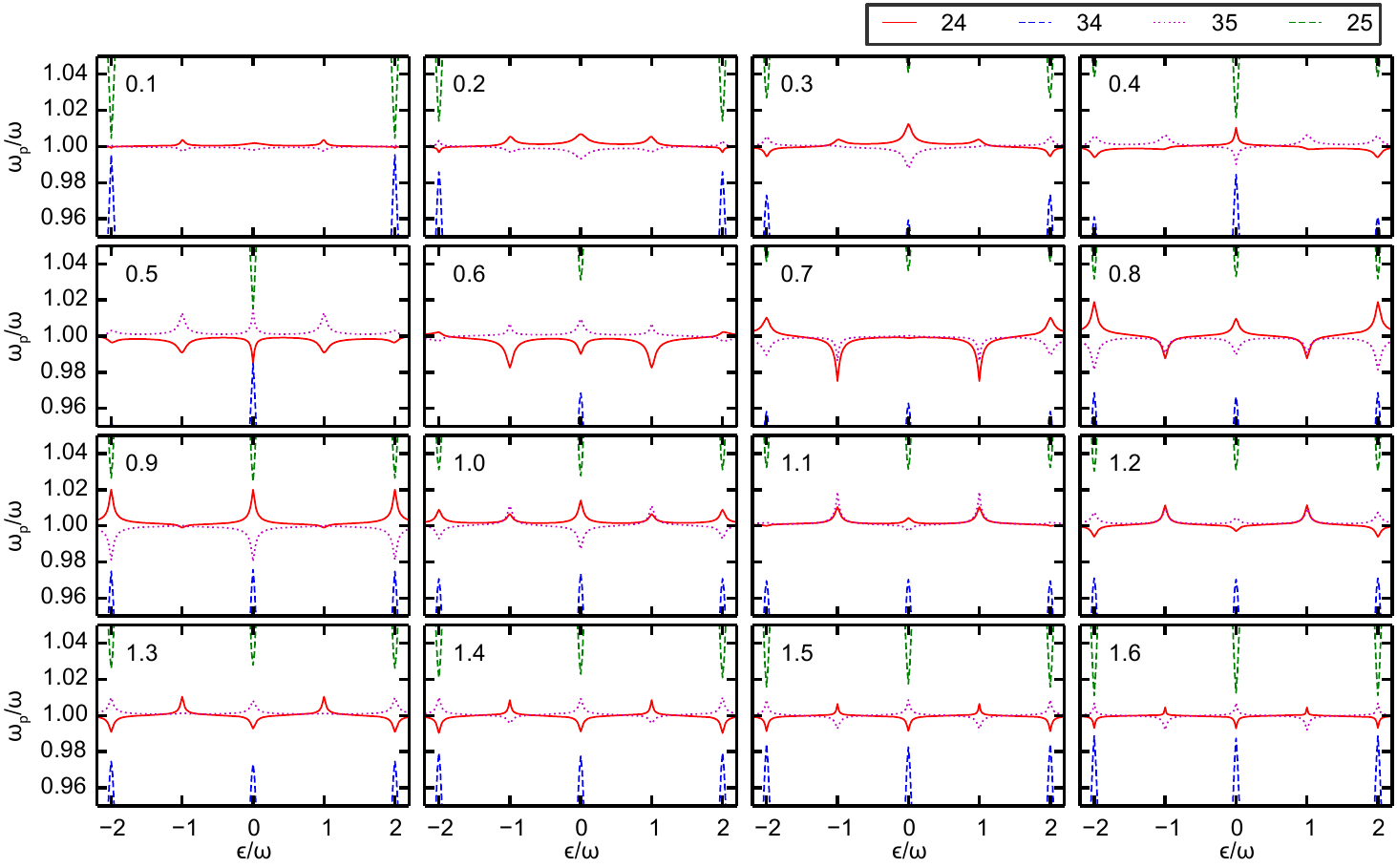}
\caption{Central frequencies of the different spectral lines involving higher energy levels as functions of qubit bias $\epsilon$, calculated from the eigenvalues of the Hamiltonian with $\Delta/\omega=0.1$. The solid red, dashed blue, dotted magenta and dashed green lines correspond to the transitions $\ket{2}\rightarrow\ket{4}$, $\ket{3}\rightarrow\ket{4}$, $\ket{3}\rightarrow\ket{5}$ and $\ket{2}\rightarrow\ket{5}$, respectively. The different panels correspond to different values of $g/\omega$, and these values are displayed inside the panels. The different panels show that the spectral lines can exhibit different patterns depending on the coupling strength. The correspondence between the coupling strength values and specific features in the spectrum is given in Table \ref{Table:SpectrumIdentificationFrom23}.}
\label{Fig:SpectraFrom23}
\end{figure*}

Noting that $E_5-E_4$ has more features than $E_1-E_0$ and $E_3-E_2$, we can expect that the transitions from the states $\ket{2}$ and $\ket{3}$ to the states $\ket{4}$ and $\ket{5}$ will exhibit a larger number of qualitative transformations (in comparison to the transitions from the states $\ket{0}$ and $\ket{1}$ to the states $\ket{2}$ and $\ket{3}$) as we increase $g/\omega$. Furthermore, Fig.~\ref{Fig:EnergyLevelSeparation} shows that the zeros and crossings of $E_3-E_2$ and $E_5-E_4$ not only are larger in number but also start at a smaller value of $g/\omega$ and end at a larger value of $g/\omega$, suggesting that utilizing the higher levels will expand the range of coupling strength values in which the qualitative-identification technique can be applied. Figure \ref{Fig:SpectraFrom23} and Table \ref{Table:SpectrumIdentificationFrom23} show that the range $0<g/\omega<\infty$ can now be divided into nine smaller intervals by observing the features at $\epsilon=0$ and $\epsilon=\pm\omega$ (with slightly finer divisions possible if we include features at $\epsilon=\pm 2\omega$).

\begin{table*}
\caption{Classification of the $\ket{2}\rightarrow\ket{4}$ and $\ket{3}\rightarrow\ket{5}$ spectral line properties according to four criteria. Each one of the sixteen values of $g/\omega$ shown in Fig.~\ref{Fig:SpectraFrom23} is assigned to one combination of features, based on the spectra shown in Fig.~\ref{Fig:SpectraFrom23} and on the matrix elements $\bra{4}(\hat{a}+\hat{a}^{\dagger})\ket{2}$ and $\bra{5}(\hat{a}+\hat{a}^{\dagger})\ket{3}$ (which cannot be immediately inferred from Fig.~\ref{Fig:SpectraFrom23}). We now have nine possible patterns that can be resolved relatively straightforwardly, compared to only five patterns that we obtained when we analyzed the shapes of the $\ket{0}\rightarrow\ket{2}$ and $\ket{1}\rightarrow\ket{3}$ spectral lines.}
\begin{tabular}{|c|c|c|c|c|c|}
\cline{3-6}
\multicolumn{2}{c|}{} & \multicolumn{2}{c|}{24 \& 35 allowed at $\epsilon=0$} & \multicolumn{2}{c|}{24 \& 35 forbidden at $\epsilon=0$} \\
\cline{3-6}
\multicolumn{2}{c|}{} & 24 peak at $\epsilon=0$ & 24 dip at $\epsilon=0$ & 24 peak at $\epsilon=0$ & 24 dip at $\epsilon=0$ \\
\hline
\multirow{2}{*}{24 peak at $\epsilon=\pm\omega$} & 35 peak at $\epsilon=\pm\omega$ & 0.3 & & 1.0,1.1 & 1.2,1.3 \\
\cline{2-6}
& 35 dip at $\epsilon=\pm\omega$ & 0.1,0.2 & & & 1.4,1.5,1.6 \\
\hline
\multirow{2}{*}{24 dip at $\epsilon=\pm\omega$} & 35 peak at $\epsilon=\pm\omega$ & 0.4 & & & 0.5,0.6 \\
\cline{2-6}
& 35 dip at $\epsilon=\pm\omega$ & & & 0.8,0.9 & 0.7 \\
\hline
\end{tabular}
\label{Table:SpectrumIdentificationFrom23}
\end{table*}

One must of course note that the spectral lines from these higher energy levels lie around the same central frequency $\omega_{\rm p}=\omega$ as the spectral lines discussed in Secs.~\ref{Sec2} and \ref{Sec3}, which could complicate attempts at using these higher energy levels in practice. On the other hand, transitions involving higher levels exhibit features up to values of $\epsilon$ that are farther away from the symmetry point than transitions among the lowest levels. Indeed, in one of the data sets in Ref.~\cite{Yoshihara} the signal from the $\ket{2}\rightarrow\ket{4}$ transition could be clearly seen between $\epsilon=\pm\omega$ and $\epsilon=\pm 2\omega$ because in that region the $\ket{0}\rightarrow\ket{2}$ and $\ket{1}\rightarrow\ket{3}$ lines do not show any significant deviation from $\omega_{\rm p}=\omega$. Therefore, in spite of some technical difficulties, it could be possible to utilize these additional spectral lines for further identification of the coupling strength.

\end{document}